\documentclass[sigconf]{acmart}

\usepackage{graphicx}

\setcopyright{none}
\acmConference[WExIR '26]{Second Workshop on Explainability in Information Retrieval, co-located with SIGIR 2026}{July 24, 2026}{Melbourne, Australia}
\acmYear{2026}
\copyrightyear{2026}
\acmISBN{}
\acmDOI{}
\renewcommand\footnotetextcopyrightpermission[1]{}

\begin{document}

\title{Interpretable Uncertainty for Adaptive Retrieval and Reasoning in Question Answering}

\author{Ritajit Dey}
\email{r.dey.1@research.gla.ac.uk}
\affiliation{%
  \institution{University of Glasgow}
  \city{Glasgow}
  \country{UK}
}

\author{Iadh Ounis}
\email{iadh.ounis@glasgow.ac.uk}
\affiliation{%
  \institution{University of Glasgow}
  \city{Glasgow}
  \country{UK}
}

\author{Graham McDonald}
\email{graham.mcdonald@glasgow.ac.uk}
\affiliation{%
  \institution{University of Glasgow}
  \city{Glasgow}
  \country{UK}
}

\renewcommand{\shortauthors}{Dey et al.}

\begin{abstract}
Large language models (LLMs) achieve a strong performance in question answering (QA), but remain prone to hallucinations and suffer from limited transparency. Retrieval-augmented generation (RAG) can improve factuality, yet decisions about when and how to retrieve from external resources are typically based on opaque policies or computationally inefficient multi-step prompting procedures. We propose an uncertainty-aware framework for adaptive QA based on explicit signals derived from LLM internal representations. We distinguish between knowledge insufficiency and knowledge ambiguity or conflict, and efficiently estimate these from hidden states in a single forward pass. These signals guide system behaviour: RAG is triggered when knowledge is insufficient, while additional reasoning is applied when ambiguity or conflict is high. By grounding adaptive decisions in decomposed and efficiently estimable uncertainty signals, this approach provides a transparent and practical alternative to existing retrieval and reasoning strategies supporting the design of interpretable user-facing tools.
\end{abstract}



\maketitle

\section{Introduction}

Question answering (QA) is a central task in information retrieval and natural language processing, requiring systems to produce accurate answers to natural language queries. Recent progress has been driven by large language models (LLMs), which generate answers directly by leveraging knowledge acquired during large-scale pretraining~\cite{petroni2019languagemodels}. Despite strong empirical performance, such models are prone to hallucinated outputs~\cite{ji2023survey}, can rely on incomplete or outdated internal knowledge~\cite{lewis2020rag}, and suffer from limited transparency in how answers are produced~\cite{bender2021stochasticparrots,wiegreffe2019attention}.

Retrieval-augmented generation (RAG)~\cite{lewis2020rag} has emerged as a dominant paradigm to address these issues by grounding generation in externally retrieved evidence. By decoupling knowledge access from answer generation, RAG-based systems can improve factuality and provide a more interpretable link between outputs and supporting information. However, retrieval is not universally beneficial: for queries where relevant knowledge is already well-represented in the LLM, retrieval may introduce noise, increase latency, and degrade answer quality~\cite{mallen2023llms}. This has motivated growing interest in adaptive retrieval strategies that determine whether and how to retrieve external evidence~\cite{asai2023selfrag,jeong2024adaptive,wu2024repoformer}.

Existing approaches to adaptive retrieval typically rely on implicit learned decision policies~\cite{asai2023selfrag,jeong2024adaptive,wu2024repoformer}, or structured multi-step reasoning-based decision mechanisms~\cite{nguyen2026hallucinate}. The former are opaque and do not disentangle different sources of uncertainty, while the latter rely on multi-step prompting and are computationally inefficient. As a consequence, it is difficult to expose interpretable, user-facing cues to end-users. This highlights the need for decision mechanisms that are both interpretable and efficient, while explicitly modelling different sources of uncertainty. In QA settings, uncertainty may arise due to knowledge insufficiency, where the model lacks the required information, or due to knowledge ambiguity or conflict, where multiple plausible interpretations or conflicting evidence exist~\cite{hullermeier2021aleatoric,kuhn2023semantic}. To illustrate knowledge insufficiency, consider the question ``Who is the head coach of Forest Green Rovers?''. Such long-tail facts about obscure teams may not be reliably represented in the model's training data; as a result, reasoning alone is unlikely to recover the correct answer (Robbie Savage), and external retrieval is required. By contrast, knowledge conflicts can be illustrated by the following question ``Who is the head coach of Chelsea?''. The training data may contain multiple historically correct answers (e.g., Thomas Tuchel, Graham Potter, Mauricio Pochettino, Enzo Maresca), each strongly associated with Chelsea Football Club. This results in competing internal representations, even though the data includes the currently correct answer (Enzo Maresca)~\cite{kandpal2023longtail}. Increasing test-time compute~\cite{snell2024scalingllmtesttimecompute} (i.e., enabling the LLM to reason via Chain-of-Thought~\cite{wei2022chainofthought} or Self-Consistency~\cite{wang2022selfconsistency}) can improve accuracy in such cases by allowing the LLM to select among competing candidate answers, as the reasoning process encourages the enumeration of relevant facts followed by the application of explicit selection criteria such as temporal recency. Hence, the aforementioned two cases call for different responses: knowledge insufficiency may benefit from external retrieval to fill gaps in the LLM's knowledge, while ambiguity or conflict may require additional reasoning or increased test-time computation to select correctly from competing candidate answers.

\begin{figure}[t]
  \centering
  \includegraphics[width=\linewidth]{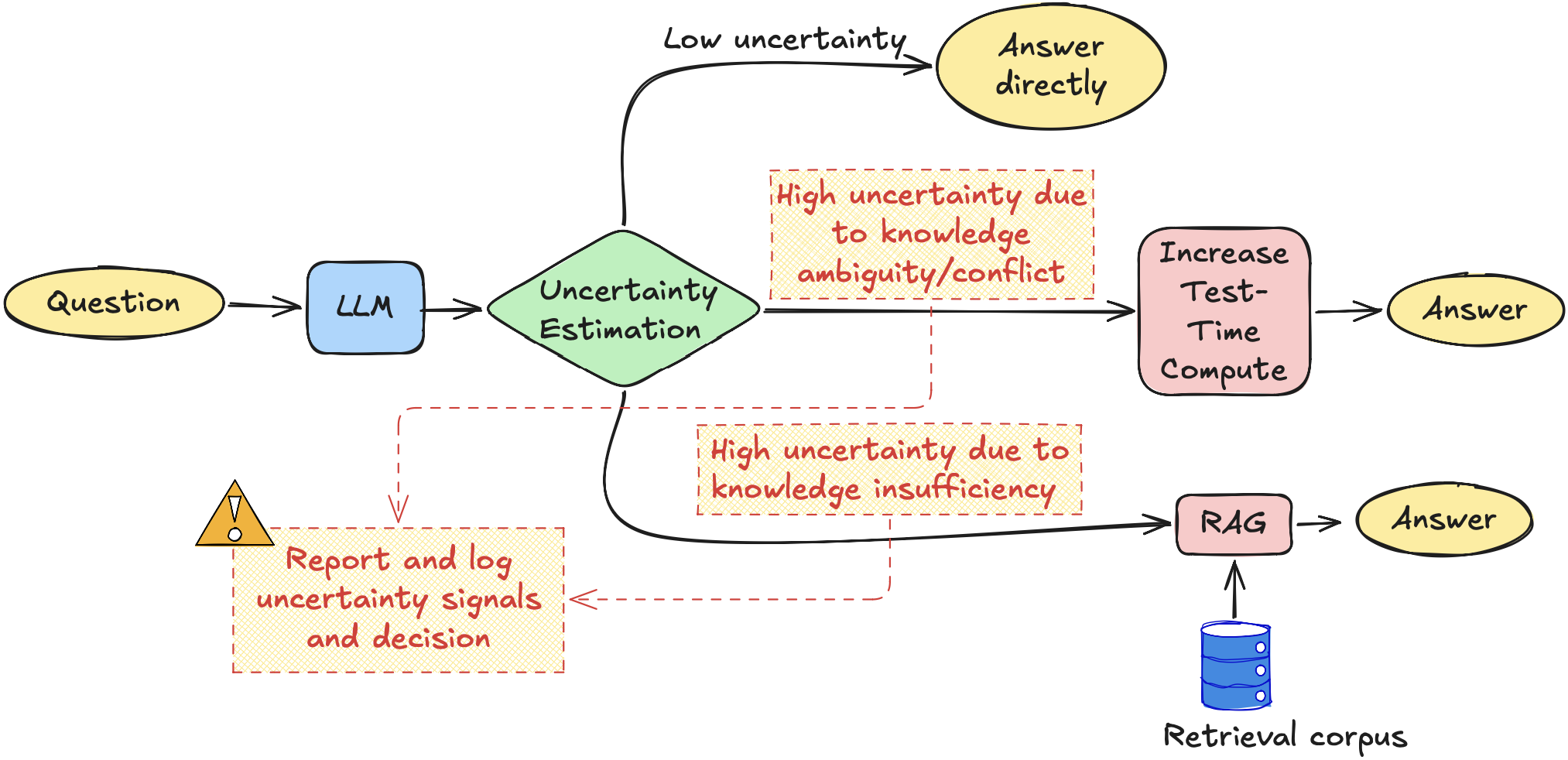}
  \caption{The proposed uncertainty-aware QA framework.}
  \label{fig:pipeline}
\end{figure}

In this work, we propose an uncertainty-aware framework for QA that explicitly distinguishes between knowledge insufficiency and knowledge ambiguity or conflict, and uses these signals to guide and explain retrieval and reasoning decisions. Our key idea is that hidden states of LLMs can be used to efficiently estimate these two forms of uncertainty within a single forward pass, avoiding the need for repeated sampling or auxiliary models. Based on these estimates, we introduce a simple decision mechanism: retrieval is triggered when knowledge insufficiency is high, while additional reasoning methods (such as Chain-of-Thought~\cite{wei2022chainofthought} or Self-Consistency~\cite{wang2022selfconsistency}) are applied when ambiguity or conflict is high. An illustration of the proposed QA pipeline is given in Figure~\ref{fig:pipeline}.

The ultimate objective is to provide a transparent and principled basis for adaptive retrieval-augmented generation in knowledge-intensive tasks, grounding routing and intervention decisions in interpretable uncertainty signals, while remaining efficient for multi-step or agentic settings where repeated sampling or prompting is impractical. These signals allow a QA system, regardless of its complexity, to surface the nature of the underlying knowledge problem at each step (e.g., missing vs. conflicting or ambiguous knowledge), and to use this information to drive behaviour as well as generate step-level explanations for why specific components (e.g., retrieval, disambiguation, clarification, abstention) are invoked. By design, this enables decision behaviour that is directly interpretable, and where needed, explainable, rather than inferred post-hoc.

\section{Methodology \& Results}

We propose an uncertainty-aware framework for adaptive QA that estimates two complementary signals from LLM hidden states: knowledge insufficiency and knowledge ambiguity or conflict, and uses them to guide retrieval and reasoning decisions.

\textbf{Uncertainty Estimation from Hidden States:} We propose estimating knowledge insufficiency and knowledge ambiguity or conflict for a given question directly from LLM hidden states using regression probes. We train lightweight regression probes over hidden states~\cite{conneau2018senteval,hewitt2019structural,hollenstein2021multilingual} to estimate two quantities:

\textbf{Knowledge insufficiency} is estimated as the number of occurrences of the fact in the pretraining corpus of the LLM, following Kandpal et al.~\cite{kandpal2023longtail} and Kang et al.~\cite{kang2023cooccurrence}, with fewer occurrences indicating higher insufficiency and uncertainty~\cite{ji2023survey}. We propose training a regression probe to predict these occurrence counts from LLM hidden states.

\textbf{Knowledge ambiguity/conflict} captures the degree to which multiple competing answers are supported by the data. For a given question, we conceptualise this quantity as the entropy of the occurrences of the different versions of the fact in the pretraining corpus, following Tomov et al.~\cite{tomov2026illusion}. Higher entropy corresponds to increased LLM uncertainty~\cite{tomov2026illusion}. We propose training a regression probe to estimate this entropy value from LLM hidden states.

Both probes operate on hidden states from a single forward pass, enabling efficient uncertainty estimation without additional sampling or prompting.

\textbf{Adaptive Decision Mechanism:} We propose using estimated uncertainty signals to control the behaviour of the QA system through a simple threshold-based decision mechanism. When a question is passed to the LLM, the hidden states of the LLM can be extracted and used to estimate knowledge insufficiency and ambiguity or conflict scores, which in turn inform downstream processing decisions. If knowledge insufficiency exceeds a given threshold, the system can trigger retrieval-augmented generation (RAG) to supplement missing information. On the other hand, when ambiguity or conflict is high, the system may instead allocate additional test-time computation~\cite{snell2024scalingllmtesttimecompute} (e.g., Chain-of-Thought~\cite{wei2022chainofthought} or Self-Consistency~\cite{wang2022selfconsistency}) to resolve competing candidate answers. When both signals are low, the model can answer directly without additional intervention.

\textbf{Preliminary Results:} We conduct experiments on factoid QA using the NQ dataset~\cite{kwiatkowski2019naturalquestions} with Llama-2-7b-chat~\cite{touvron2023llama2}, using Semantic Entropy (SE)~\cite{kuhn2023semantic} and Weighted Entropy Production Rate (WEPR)~\cite{moslonka2026hallucinationdetection} to selectively trigger RAG. Results show that our framework significantly outperforms (according to McNemar's test) both the LLM-only baseline (+5.9\% for SE, +4.7\% for WEPR) and the always-on RAG baseline (+3.3\% for SE, +2.1\% for WEPR). This supports the promise of uncertainty-driven adaptive QA.

\section{Conclusions}

We advocated a shift in adaptive question answering (QA) from opaque policies toward decisions grounded in explicit, interpretable uncertainty. By distinguishing between knowledge insufficiency and knowledge ambiguity or conflict, we outlined a framework where different forms of uncertainty drive different actions -- retrieval to acquire missing information, and additional reasoning to resolve competing alternatives. This decomposition enables interpretable decision-making by exposing the type of knowledge problem the system is facing and justifying its behaviour accordingly, thereby providing both users and system designers with step-level explanations, as needed. We further argued that interpretability should not come at the cost of efficiency. Our approach highlights the potential to estimate these signals within a single forward pass, avoiding expensive sampling or auxiliary components, which is critical for deployment in complex settings. More broadly, we advocated for treating interpretability as a design principle rather than a post-hoc tool in QA systems. This allows users to understand why the system searched or reasoned further, and enables system designers to audit and diagnose behaviour in terms of missing or conflicting knowledge.

\bibliographystyle{ACM-Reference-Format}
\bibliography{references}

\end{document}